# Electrically functionalized body surface for deep-tissue bioelectrical recording

Dehui Zhang[1], Yucheng Zhang[2], Dong Xu[2], Shaolei Wang[3], Kaidong Wang[3], Boxuan Zhou[2], Yansong Ling[2], Yang Liu[2], Qingyu Cui[3], Junyi Yin[4], Enbo Zhu[3], Xun Zhao[4], Chengzhang Wan[2], Jun Chen[4], Tzung K. Hsiai[3, 4], Yu Huang[2,5,*], and Xiangfeng Duan[1,5,*]

[1] Department of Chemistry and Biochemistry, University of California, Los Angeles, Los Angeles, CA, USA.
[2] Department of Materials Science and Engineering, University of California, Los Angeles, Los Angeles, CA, USA.
[3] Division of Cardiology, Department of Medicine, School of Medicine, University of California, Los Angeles, Los Angeles, CA, USA.
[4] Department of Bioengineering, University of California, Los Angeles, Los Angeles, CA 90095, USA.
[5] California NanoSystems Institute (CNSI), University of California, Los Angeles, Los Angeles, CA, USA.
*Emails: xduan@chem.ucla.edu (X. D.), yhuang@seas.ucla.edu (Y. H.)

**Directly probing deep tissue activities from body surfaces offers a noninvasive approach to monitoring essential physiological processes[1-3]. However, this method is technically challenged by rapid signal attenuation toward the body surface and confounding motion artifacts[4-6] primarily due to excessive contact impedance and mechanical mismatch with conventional electrodes. Herein, by formulating and directly spray coating biocompatible two-dimensional nanosheet ink onto the human body under ambient conditions, we create microscopically conformal and adaptive van der Waals thin films (VDWTFs) that seamlessly merge with non-Euclidean, hairy, and dynamically evolving body surfaces. Unlike traditional deposition methods, which often struggle with conformality and adaptability while retaining high electronic performance, this gentle process enables the formation of high-performance VDWTFs directly on the body surface under bio-friendly conditions, making it ideal for biological applications. This results in low-impedance electrically functionalized body surfaces (EFBS), enabling highly robust monitoring of biopotential and bioimpedance modulations associated with deep-tissue activities, such as blood circulation, muscle movements, and brain activities. Compared to commercial solutions, our VDWTF-EFBS exhibits nearly two-orders of magnitude lower contact impedance and substantially reduces the extrinsic motion artifacts, enabling reliable extraction of bioelectrical signals from irregular surfaces, such as unshaved human scalps. This advancement defines a technology for continuous, noninvasive monitoring of deep-tissue activities during routine body movements.**

Deep tissue activities, such as body blood circulation, brain functioning, and muscle movements, offer vital insight into essential physiological processes, including cardiovascular function, neural activity, and muscular coordination. Such activities can manifest as body surface electrical potential[1] and impedance[2,3] changes, although they often suffer from substantial attenuation and perturbation after traversing various tissue and body surface contact layers. Establishing reliable electrical contacts with the body surface represents a critical technical challenge due to substantial topographic, mechanical, and electrical mismatch between the body surface and the typical electrical readout circuit[4,5]. In particular, human body surfaces feature highly irregular, non-Euclidean surface geometry on both macroscopic (e.g., knuckles) and microscopic (such as wrinkles and unshaved skin/scalp) scales. Additionally, the soft and deformable nature of human tissues features a large mechanical mismatch with rigid electrodes, resulting in substantial contact impedance fluctuations during routine body movements that could further degrade the deep-tissue signals[6]. Furthermore, to ensure high-fidelity monitoring of intrinsic physiological signals, it is essential that *the physiological deformation of soft tissues is neither impeded by the typically stiff electronic probes nor impairs their functionalities*. **These complexities pose fundamental challenges in using conventional planar electronics to establish stable, low-impedance bioelectrical contacts, which are essential for accurately capturing the minuscule signals emanating from the body surface.**

The electrical signal probed from body surface is fundamentally governed by two field distributions, i.e., the electromagnetic field $E(x,y,z)$ following the Maxwell equations, and the mechanical strain distribution $\varepsilon(x,y,z)$ following the Newtonian mechanics (Fig. 1a). Specifically, $\varepsilon(x,y,z)$ modulates $E(x,y,z)$ and the impedance distribution $\rho(x,y,z)$ through geometrical deformation, biofluid redistribution, and strain-induced impedance change. Consequently, the current measured in the contact area ($A$), $I_{contact} = \oiint E(x,y,z)/\rho(x,y,z)dA$, reflects all tissue mechanical changes from the contact to the deep tissues within the conductive paths. The contact artifacts perturb the deep tissue signal by introducing $\Delta\rho_1$ in strained tissue beneath the contact, $\Delta\rho_2$ at the skin-electrode interface, and $\Delta\rho_3$ inside the contact material layer (Fig. 1b), and thus also redistributing $E(x,y,z)$. High $d\rho_{1,2,3}/d\varepsilon$ values will make it formidable to resolve the deep tissue signals from the body surface electrodes, which extrinsically vary over time due to body motion and contact adhesion/conductance degradation.

The current commercial electrodes[7,8] use ionic gel conductors to ensure a conformal body electrical interface. However, it often suffers from large $\Delta\rho_1$, $\Delta\rho_2$, and $\Delta\rho_3$ due to large mechanical mismatch **and inadvertently traps air bubbles** (at the contact interface) that could vary substantially under body surface strain. This approach is thus highly susceptible to motion artifacts and thus is typically only applicable in a highly controlled rest state in specialized medical offices. Additionally, **the high contact impedance typically necessitates relatively large-area pads (>0.5 cm diameter)**, which leads to substantial spatial signal averaging and limits the spatial resolution. Furthermore, the large-area pads could cause additional problems such as low breathability, high surface strain, and undesirable irritation[9], which limit the comfort level and could cause skin rash, particularly for more delicate subjects, including infants[10,11].

Some recent advances[12-22] implement new contact materials, such as metal nanomeshes[9,11,12,23] and 2D materials[13,24,25], to assure soft and conformal skin-electronic interfaces and mitigate $\Delta\rho_1$ and $\Delta\rho_2$. Specifically, microscale conformal nanomaterials were applied onto the skin without any mechanical support from polymer films, resulting in greatly improved conformability and breathability[13,15,26]. However, on-skin metal nanowire meshes transport electrons through point contacts between different nanowires. Such point contacts are inherently sensitive to nanoscale mechanical detachment and reconnections, showing strong strain-induced impedance change $\Delta\rho_3$, and are more suitable for tracking surface deformation than the intrinsic electrophysiological signals[13,16]. On the other hand, van der Waals thin films (VDWTFs) formed among the staggered 2D nanosheets feature broad-area sliding vdW interfaces between dangling-bond-free nanosheets (Fig. 1c, d) and can retain excellent electronic interface even under excessive external strain (low $\Delta\rho_3$), **offering an appealing materials system for the applications that demand minimal beneath-contact ($\Delta\rho_1$), at-contact ($\Delta\rho_2$), and in-contact ($\Delta\rho_3$) impedance change during body motion, desirable for probing deep tissue electrophysiological signals.**

Nonetheless, **traditional electrode preparation methods often require high vacuum or high temperature steps to achieve high-quality electrodes, making direct application onto body surfaces challenging. Typically, these electrodes are prepared separately and transferred onto the body surface using polymer membranes, which often struggle with conformality and adaptability while maintaining a roust direct electronic interface with dynamic body surfaces.** Herein, by formulating and directly spray coating biocompatible 2D nanosheet ink onto the human body under bio-friendly conditions, we create microscopically conformal and adaptive VDWTFs that seamlessly merge with non-Euclidean, hairy, and dynamically evolving body surfaces. This gentle process allows high-performance VDWTFs to form directly on body surface, resulting in electrically functionalized body surfaces (EFBS). The EFBS features nearly two orders of magnitude lower contact impedance than the current gel electrodes and substantially suppresses motion artifacts to enable vastly improved signal extraction. It allows to reliably probe biopotential and bioimpedance at different body areas, including wrist, neck, and unshaved scalp, and monitor a large variety of deep-tissue activities, such as artery contractions, muscle movements, saliva secretion, brain blood flow, and brain electrical potentials with greatly improved signal-noise ratio (SNR) (15.9 dB in motor test and 8.3 dB in radial artery impedance, see detailed comparisons in Supplementary Table 1). The EFBS provides a unique approach for robust interrogation of deep tissue activities under routine body movements.

**Conformal and adaptive VDWTF-EFBS**

The biocompatible 2D-ink consisting of the 2D $MoS_2$ nanosheets formulated in isopropyl alcohol (IPA)[27,28]. IPA is routinely used for skin sterilization. A spray-coating process was then used to form microscopically conformal van der Waals thin films (VDWTFs) directly on irregular, dynamically evolving body surfaces under ambient conditions (Fig. 1f), producing electrically functionalized body surfaces (EFBS). **The directly spray-coated VDWTF-EFBS offer several critical features to ensure a robust bio-friendly low-impedance interface for probing deep tissue signals: (1)** $MoS_2$ is a widely used solid lubricant and is biodegradable without noticeable biohazard at practical doses[29,30]. No perceptible irritation or allergic reaction was observed during the prolonged usage (weeks) in the tests for four subjects. **(2)** The atomically thin 2D nanosheets are extremely flexible and can actively adapt to microscopic skin surface texture during the spray-coating process, forming conformal interfaces that ensure **reliable physical contact**. **(3)** The dangling-bonding-free basal planes of the $MoS_2$ endow nanoscale hydrophobicity, ensuring **a chemically compatible interface and robust nanoscale adhesion** to the lipid bilayers in the stratum corneum on the skin surface[31,32] (Fig. 1e), which has been proven critical for further reducing skin-contact impedance in other material systems[33,34]. **(4)** The spray-coated VDWTFs feature broad-area vdW interfaces among the staggered nanosheets that can slide laterally against each other (Fig. 1c, d). This leads to **exceptional stretchability** and minimal mechanical-induced impedance change as compared to other skin coatings with 1D or 0D building units that feature delicate point contacts between different crystals, maximizing both the mechanical stretchability and electrical stability. **(5)** The winding nanochannels around the edges of the staggered nanosheets provide **breathability** (see Fig. 2n in ref. 25), which is crucial for minimizing potential skin irritations and mitigating sweat-induced delamination, further ensuring a robust mechanical and electrical interface during prolonged applications.

Fig. 1e shows an equivalent circuit of the skin-electrode interface. The total contact impedance $Z_{total}$ depends on the series impedance components. In particular, the contact-skin interface impedance $Z_I = R_I/(1 + j\omega R_I C_I)$ is the carrier transport bottleneck, and its susceptibility to tissue deformations ($dZ_I/d\varepsilon$, where $\varepsilon$ is the strain, corresponding to $\Delta\rho_2$) is also a major contributor to motion artifacts. The strong adhesion of the conformal skin-coating eliminates any micro/nanoscale contact gaps (air bubbles), minimizing $Z_I$ and $\Delta\rho_2$ (Supplementary Fig. 1-3). The sliding 2D interfaces inside the VDWTF further suppress the impact of surface deformations on lateral resistance within VDWTF (minimize $dR_L/d\varepsilon$, corresponding to $\Delta\rho_3$). A numerical simulation of the micro/nanoscale mechanics is included in Methods.

The directly spray-coated 2D VDWTFs feature seamless skin-conformal and adaptable interfaces (Fig. 1g, h), thus transforming arbitrary body surfaces into low-impedance, motion-artifact-free electrical interfaces. Depending on the exact body surface area, we used different readout electrodes to extract the signal from the VDWTF-EFBS, including gold-SEBS flexible electrodes for relatively flat, regularly moving skin surfaces for further strain relaxation (Fig. 1i), or commercial rubber-packaged electrodes for EEG measurement on unshaved scalps (Fig. 1j) for better benchmarking against the state-of-the-art results. For a robust interface, the readout electrodes were first coated with a thin layer of graphite and NaCl-doped conductive syrup. The graphite syrup sticks well to a large variety of surfaces, including $MoS_2$ and metal electrodes. The viscosity of graphite syrup was controlled by tailoring graphite and water concentration so that it fully releases the contact strain (zero Young's modulus) with the underlying EFBS. In this way, the VDWTF-EFBS can be implemented for monitoring a variety of deep-tissue activities, including electrical potentials generated by the body (ECG and EEG), or impedance change of tissues (artery, muscles, biofluids in the scalp, etc.) (Fig. 1k).

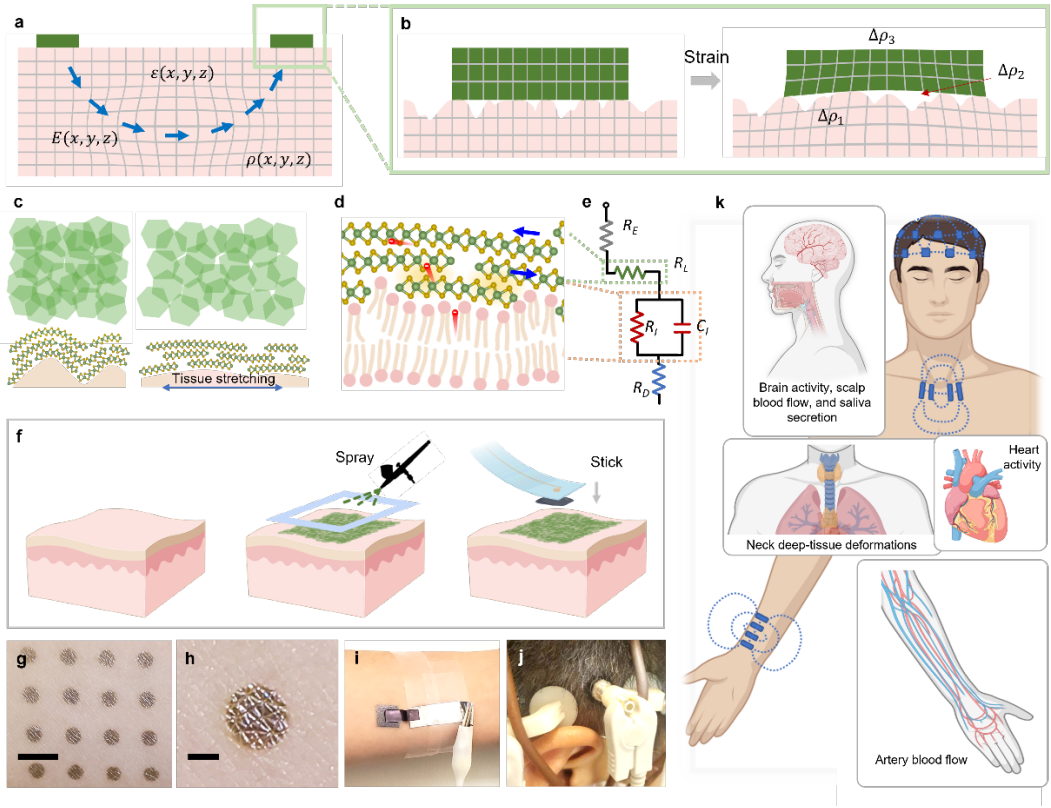

**Fig. 1| Stretchable and adaptative electrically functionalized body surfaces for deep tissue sensing.**
**a**, A schematic of electric field and strain distribution when the deep tissue activity (here illustrated as an internal strain) is probed with body surface electrodes. **b**, When the contact is strained, additional impedance change may happen beneath the contact in tissue ($\Delta\rho_1$) due to Young's modulus mismatch, at the contact ($\Delta\rho_2$) due to partial interface detachments, and in the electrode ($\Delta\rho_3$) due to electrode material distortion. These impedance changes are extrinsic to the deep tissue activities and should be avoided. Specifically, commercial gel electrodes and some dry electrodes are plagued by trapped air bubbles or airgaps at the non-planar body surface (white gaps in the middle), leading to larger $\rho_2$ and $\Delta\rho_2$. **c**, Top view (upper) and side view (lower) of VDWTF forming conformal interfaces with the textured body surface. **d**, The zoom-in side view of **c**. The 2D nanosheets slide at the nanoscale and minimize the carrier transport degradation during tissue morphology change, providing higher resilience to motion artifacts than other skin-conformal coatings. Moreover, the hydrophobicity of the dangling-bond-free 2D basal plane enables seamless adhesion between the nanosheets and the skin lipid bilayers to ensure low contact impedance. **e**, The equivalent circuit of the contact impedance, including the electrode series resistance ($R_E$), the lateral resistance inside the VDWTF ($R_L$), the MoS$_2$-skin interface resistance ($R_I$) and capacitance ($C_I$), and the deeper tissue resistance ($R_D$). **f**, The 'spray-and-stick' process for implementing the VDWTF-EFBS. MoS$_2$ nanosheets dispersed in IPA were directly sprayed on the body surface through a stenciled mask (light blue in the figure). The IPA quickly evaporates, leaving the MoS$_2$ as a body-conformal conductive thin film. Next, metal electrodes pre-coated with a thin conductive adhesive (graphite-doped syrup) were attached onto the VDWTF-EFBS. **g**, A VDWTF-EFBS array patterned with a stencil mask. Skin textures are clearly visible with even enhanced contrast due to stronger reflectivity. Scale bar: 5 mm. **h**, Zoomed-in view of g. Due to the transient ink flow before IPA evaporation, the 2D film conforms over the body surface, including the trenches that conventional electrodes fail to fully cover. Scale bar: 1 mm. **i**, Connection with solid metal electrodes. A flexible gold-on-SEBS film (left) bridges the VDWTF-EFBS with aluminum tapes to further release the strain. The package is applied to VDWTF-EFBS in areas with less hair, including the arms and the neck. **j**, EFBS connected to a commercial headset for unshaved scalp EEG tests, in which the electrodes are pinched to the scalp. **k**, A schematic illustrating the VDWTF-EFBS probing different deep-tissue activities, including electrical potentials generated by the body (ECG and EEG), or impedance change of tissues (artery, muscles, biofluids in the scalp, etc.) during bioactivities.

**VDWTF-EFBS with reduced contact impedance and suppressed motion artifacts**

We first performed electrical impedance spectroscopy (EIS)[35] with our VDWTF-EFBS on the forearm (Fig. 2a), revealing a 70 times lower DC and AC contact impedance than commercial Ag/AgCl gel electrodes (Fig. 2b). The DC contact resistance is $10.8\pm3.2$ k$\Omega \cdot$cm$^2$ measured in four separate implementations, notably lower than any wet/dry body surface electrodes reported to date (Supplementary Table 2). Furthermore, the VDWTF-EFBS showed no degradation in contact with water (Fig. 2c) and with extended water rinse, demonstrating its robustness in wet environments. Next, we tested the lateral resistance of VDWTF-EFBS on the knuckles with repeated bending (Fig. 2d, e). The knuckle region is chosen for its complex surface geometry with large mechanical change during bending. Notably, the VDWTF-EFBS across the knuckles show only ~20% current decrease when bent (Fig. 2f, g), even with the contact artifacts counted in. **This is fundamentally different from the same measurements with metal nanomeshes[16], nanowire networks[13], and in-situ printed electrodes[36], where the resistance increases by >300% when bent.** Furthermore, the lateral conductance does not significantly degrade under 100 repeated finger bending cycles despite the drastic surface geometry change.

Prior studies typically compare the motion artifacts of different bioelectronic devices by measuring the signals under body motions[37,38]. The method triggers both intrinsic tissue potential/impedance change and contact artifacts from body surface-electrode mechanical interfaces. We have adopted an alternative measurement setup for better separating the intrinsic bio-electromechanical signals from the extrinsic contact mechanical artifacts (Fig. 2h). Two electrodes were attached to the forearm to measure the body resistance. An eccentric motor introduces external vibration either to the body surface between two electrodes (Fig. 2n), or to the electrical wires (Fig. 2q). Unlike the typical tests with the muscle and skin stretching in a much larger area, the locally triggered vibration resides near the motor and attenuates over longer distances. Hence it better separates the intrinsic bioimpedance change from the extrinsic contact impedance change.

We compared the performance of commercial Ag/AgCl gel electrodes (middle column, Figs. 2i, l, o, and r) and the VDWTF-EFBS (right column, Figs. 2j, m, p, and s), with Figs. 2k, n, and q denoting the motor arrangement of plots in the same row. Figs. 2l, m show slow natural fluctuations of arm resistance at rest. The on-skin motor introduces fast resistance fluctuations from tissue movement that are similar for both contacts (Fig. 2o, p). On the other hand, the on-wire vibration significantly increases the noise for the commercial Ag/AgCl gel electrodes (Fig. 2r) and reduces its SNR by 21.6 dB after the motor is turned on. In comparison, the SNR decreases by only 5.7 dB with VDWTF-EFBS (Fig. 2s and Supplementary Fig. 4) (see Methods for SNR evaluation), demonstrating a substantial suppression of the extrinsic motion artifacts at body surface-electrode interfaces.

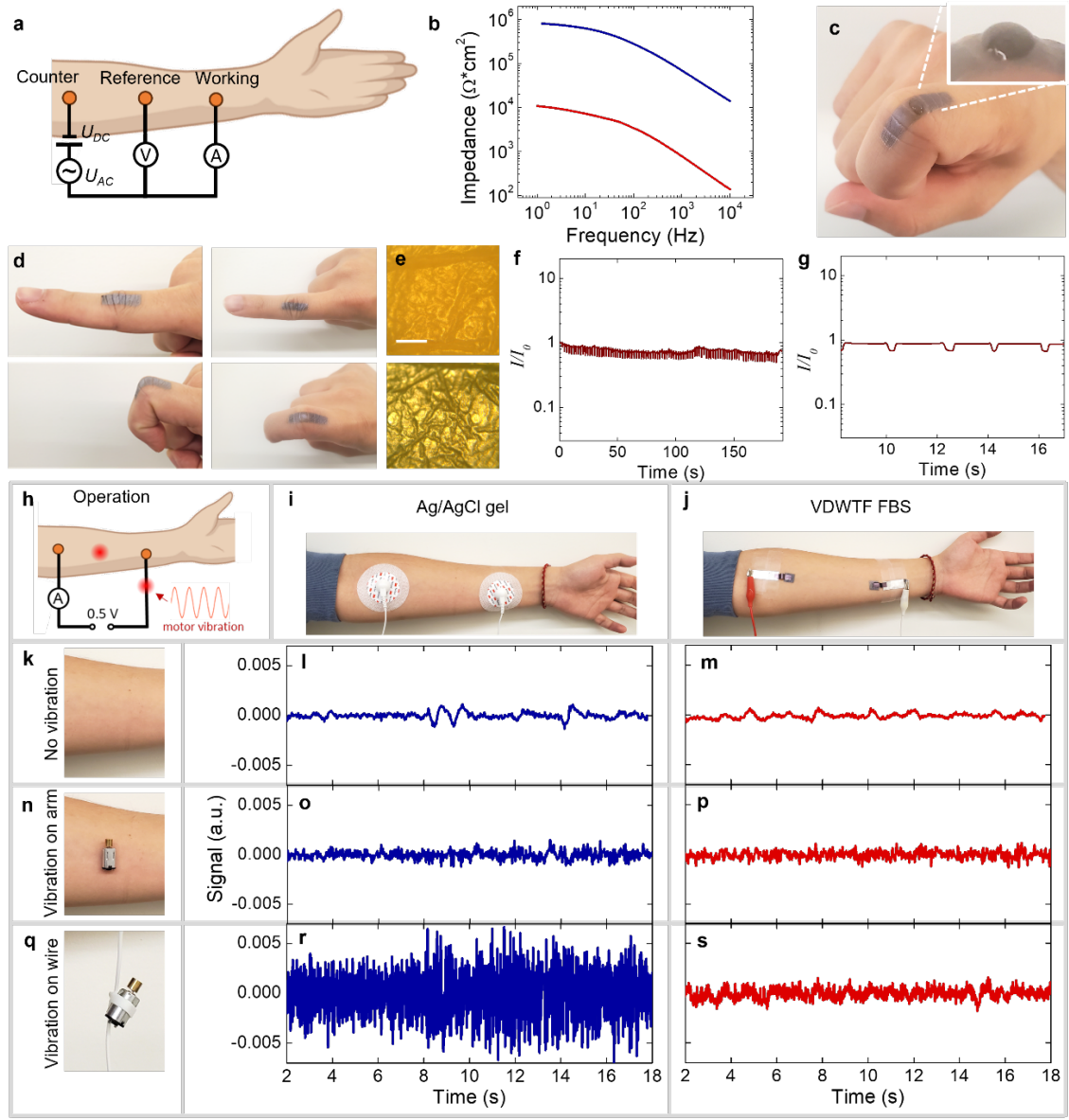

**Fig. 2| Basic characterizations of VDWTF-EFBS. a**, The EIS measurement setup. **b**, Impedance measured on the forearm contacted with VDWTF-EFBS (red) and commercial Ag/AgCl-gel pads (blue). The lowest frequency (corresponding to the right tips of the curves) is 1 Hz. **c**, A water droplet on the EFBS. **d**, VDWTF-EFBS on knuckles for the in-plane resistance test. **e**, Microscope photo of bent knuckle surface before (upper) and after (lower) applying the EFBS showing similar skin texture. The optical contrast is low for bare skin due to its semitransparency. Scale bar: 500 $\mu$m. **f**, Measured current through the VDWTF-EFBS over 100 repeated knuckle-bending, showing little conductance degradation. **g**, The zoomed-in waveform shows only ~20% resistance increase when bent. **h**, The DC resistance measurement setup with different electrodes. A 0.5 V bias was applied between two terminals along the inner side of the left forearm. Eccentric motors were applied to the red spot regions (on the skin and on the wire) to generate local mechanical oscillations separable from body motions. **i**, **j**, **k**, **n**, and **q** are the test layout for the corresponding row/column. Commercial gel electrodes (**i**) and VDWTF-EFBS (**j**) were used to collect the corresponding columns of data. Different rows represent results without motor oscillation (**k**), with oscillation on the arm (**n**) and on the wire (**q**). **l**, **m**, **o**, **p**, **r**, and **s** are the corresponding currents. With extrinsic wire oscillation, the VDWTF-EFBS electrodes **s** show notably better immunity to the motion artifacts than the commercial Ag/AgCl gel electrodes **r**.

**Bioimpedance changes on the radial artery**

With the greatly suppressed contact artifacts demonstrated above, we used the VDWTF-EFBS to measure the bioimpedance changes on the radial artery. Blood is rich in ions, forming a preferred charge transport pathway compared to other tissues, such as muscles, fat, and bones[39]. During the blood flow, the artery diameter contracts and expands due to the pulsed blood pressure change (Fig. 3a), modulating the artery impedance. We applied a four-probe AC impedance measurement above the radial artery on the wrist to inject a 90 kHz AC current (see Methods) through contacts 1 and 4, and measured the voltage drop between contacts 2 and 3 with a lock-in amplifier (Fig. 3b). AC signal capacitively couples the electrodes to the tissue beneath the skin, reducing the high body surface impedance. With a constant AC current between contacts 2 and 3, the AC voltage drop between contacts 2 and 3 directly reflects the impedance. The AC four-probe impedance measurement is well-adopted in electrical impedance tomography (EIT) for evaluating deep-tissue activities[40,41].

Fig. 3c shows the measured voltage $U$ (red) synchronized with the electrocardiography (ECG) signals. The systolic (Fig. 3d, pink) and diastolic (light blue) sections are highlighted for one pulse cycle. During the pulse cycle, the inflow of blood first expands the measured artery region and decreases the impedance (S to P point in the plot), leading to a lower voltage drop between the inner electrodes. The outflow of blood then gradually decreases the artery diameter and increases the impedance, followed by a diastolic peak (D) formed by the reflected wave that occurs from the blood crash into the aortic valve by the blood pressure of the aorta.

The waveform shows notably suppressed noise and 8.3-dB higher SNR than commercial Ag/AgCl gel pad results (Fig. 3c and Supplementary Fig. 5). Moreover, we observed narrower P peaks with the VDWTF-EFBS (peak width = $(0.179\pm0.028)T$, $T$ is the pulse rate) than with commercial pads (peak width = $(0.254\pm0.046)T$) (see Methods for evaluation details). The difference is attributed to the narrower EFBS strips (5 mm for both stripes and spacings) than the commercial pads (circular, 10 mm in diameter and inter-electrode spacings). The commercial pads need larger areas to achieve stable adhesion and low enough surface contact impedance, which leads to signal average over the larger area, lowering the spatial resolution and blurring the observed waveform. Better-resolved pulse waveforms are essential for extracting cardiovascular health information, as suggested in previous works[42,43]. In particular, the AC measurement of arterial pulse wave is extremely sensitive to contact artifacts. The pulse wave typically only induces 0.1~0.3% fluctuations of the total impedance between the inner two electrodes, while surface contact artifacts could easily produce >1% signal fluctuation, which could completely mask up the signal of interest. Consequently, the VDWTF-EFBS as an ultra-low-noise bioelectrical interface is essential for deep-tissue sensing.

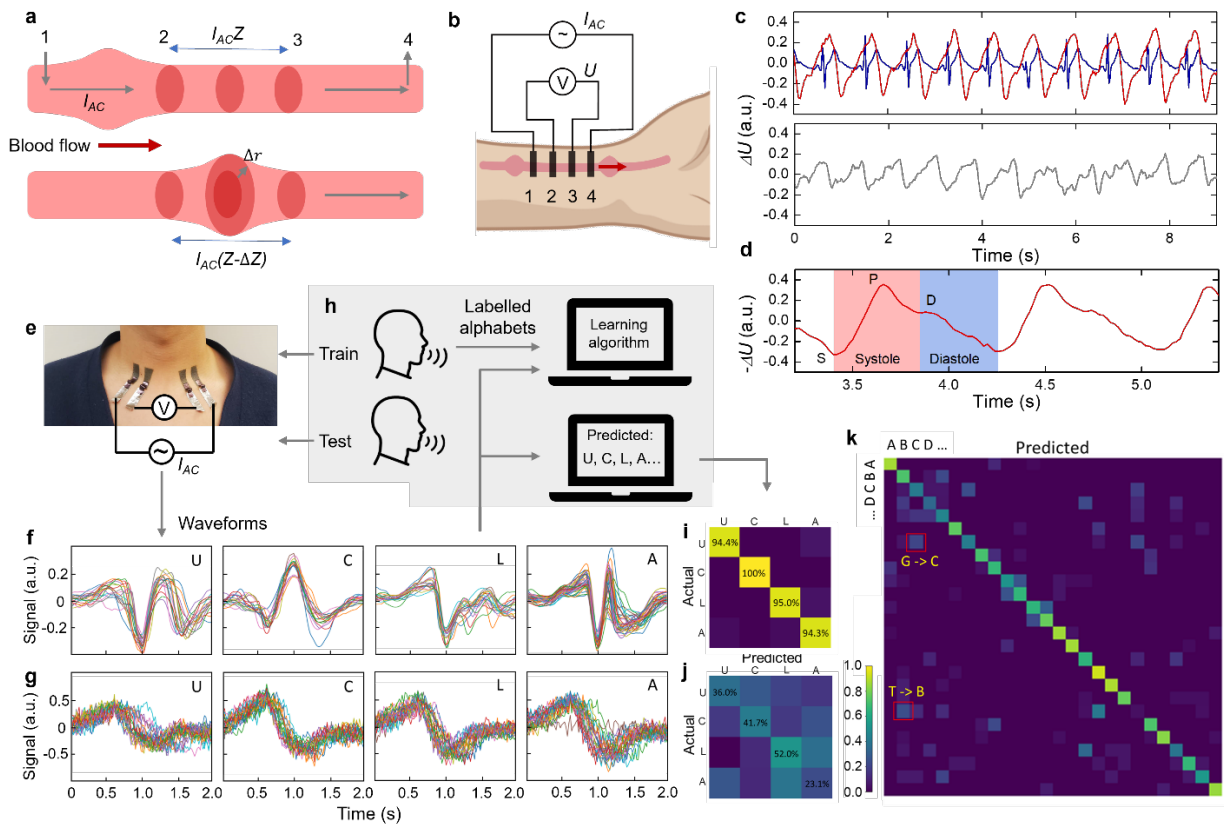

**Fig. 3| The VDWTF-EFBS to probe deep tissue impedance changes. a**, A diagram illustrating the mechanism for measuring the arterial pulse waves using the impedance tracking method. Pulsatile blood flow induces a circumferential stretch to the arterial wall, causing an impedance reduction $\Delta Z$, which changes the voltage between contact 2 and 3 in the four-probe measurement illustrated in **b**. **b**, Implementation of the VDWTF-EFBS on the wrist. The four EFBS contacts ('1', '2', '3', and '4') are patterned on the wrist area above the left radial artery (rendered in red). A 90-kHz sinusoidal current of 12 $\mu$A is injected into the arm through contacts 1 and 4. The AC voltage between contacts 2 and 3 is captured with a lock-in amplifier. **c**, Upper: The arterial pulse waveform (red) measured on the wrist, with the ECG waveform (blue) recorded between the two forearms. Lower: waveforms recorded by commercial gel electrodes later on the same wrist. **d**, The zoomed-in plot of **c** within two pulses. A minus sign is added to $\Delta U$ for better correlation to the pressure change. **e**, A photo of the VDWTF-EFBS applied to the neck region. The rigid electrodes and the connection interfaces are placed on the clavicles, while the VDWTF-EFBS extends to the regularly moving neck region to probe the underneath deep-tissue impedance change. **f**, The example waveforms measured during the repeated pronunciation of the alphabets, i.e., 'U', 'C', 'L', and 'A', showing distinct and repeatable patterns. **g**, The waveforms measured with commercial Ag/AgCl gel electrodes at the same position, showing non-distinguishable patterns with different alphabets. **h**, The training and test scheme of machine learning algorithms for retrieving the alphabets from measured waveforms. In the training session, the waveforms are labeled with the corresponding alphabets to train the algorithm. In the test session, only raw waveforms are inputted to the computer. The output predicted alphabets are compared with the ground truth to quantify the classification accuracy. The arrows represent the information flow. (**I**) Vocal classification based on the measured waveforms in **f**. The vertical axis labels the actual letters spoken, while the horizontal axis labels the classification results. Different colors represent the prediction probability. The nearest centroid classifier gives an overall classification accuracy of 96%. **j**, The classification results for commercial Ag/AgCl gel electrodes corresponding to **g**. **k**, Classification of all 26 alphabets. The diagonal pattern indicates a general correlation between the waveforms and the alphabets.

**Bioimpedance change in the neck during phonation**

Biomechanical deformation of other deep tissues also introduces bioimpedance variations. Many important bioactivities, such as breath, swallowing, neck movements, and phonation, deform the tissue lying beneath the neck, making the region of particular interest. Specifically, we probe the tissue impedance during phonation by deploying the VDWTF-EFBS between the two clavicles (Fig. 3e). The VDWTF-EFBS extends to the neck region to collect the signals from the soft, dynamically moving neck regions, and causes no mechanical strain or discomfort on the neck. The inner two EFBS electrodes are located at the inner tips of the clavicles, with 8-mm stripe width and spacings to the outer stripes. The subject pronounces the 26 alphabets, each repeated 25 times over 3 hours. Fig. 3f shows the example waveforms measured during the phonation ('U', 'C', 'L', and 'A'), with repeatable distinct features for different letters. We also performed a control measurement with the commercial pads directly attached to the same neck region. The waveforms are notably noisier and show no apparently distinguishable features during repeated pronunciations of different letters (Fig. 3g), suggesting the collected signals are largely dictated by contact artifacts from the rigid electrode interfaces. The comparison further confirms the necessity of using VDWTF-EFBS for robustly capturing deep neck tissue movements.

We further retrieve the spoken alphabet from the waveforms with a nearest centroid classification algorithm (see Methods). We first train the algorithm with randomly selected example alphabet-waveform pairs, and then test the model's prediction accuracy with new waveforms (Fig. 3h). **The four-alphabet classification (Fig. 3i) accuracy is 96%. In comparison, commercial Ag/AgCl gel electrodes give a much lower accuracy of 38% (Fig. 3j)**. Fig. 3k plots the classification result over all 26 alphabets, suggesting a clear, universal correlation between the waveforms and the vocal activity. However, some alphabets tend to mix, such as 'G' and 'C', with 21.1% of the pronounced 'G' waveforms misidentified as 'C', and 23.5% of 'C' misidentified as 'G'. The mixing is attributed to the similar neck tissue movements during the phonation, as the major vocal difference between them was generated at the front region in the oral cavity, such as teeth and the tongue tip. The difference is less correlated to the deep-neck impedance and, hence, is only modestly separable in our measurements.

**VDWTF-EFBS on unshaved scalps**

Another advantage of the directly spray-coated VDWTF-EFBS interface is its compatibility with highly irregular body surface geometries. The hairs of different individuals have very different surface profiles, causing additional challenges in forming good electrical interfaces. In this context, the directly spray-coated VDWTF-EFBS can be readily applied on unshaved human scalps and offers essential immunity to such irregularities. The side view of cowhide with spray-coated VDWTFs (Fig. 4a) shows that the VDWTF-EFBS can readily penetrate deep in the hair to form a continuous conformal film on the underlying surface (skin). We connected the VDWTF-functionalized hairy scalp to a commercial EEG headset with the standard 10-20 system electrode layout (Supplementary Fig. 6) and recorded the electrical potential waveforms with the eyes closed (Fig. 4b). Characteristic alpha (8-12 Hz) and beta (peaks around 20 Hz) waves were observed in the Fourier-transformed spectra (Fig. 4c), which showed frequency patterns similar to that measured using recommended standard conductive gel on the same EEG headset (Supplementary

Fig. 7a-c), validating the efficiency of the VDWTF-EFBS. Furthermore, the VDWTF-EFBS provides 7.3-dB higher SNR and 2.2-dB higher signal than commercial electrodes as evaluated at the alpha-wave hotspot at P3-A1 (blue boxes in Fig. 4c and Supplementary Fig. 7b). The alpha wave is predominantly observed in the occipital lobes (Fig. 4d) and suppressed when eyes are open (Supplementary Fig. 7d-f).

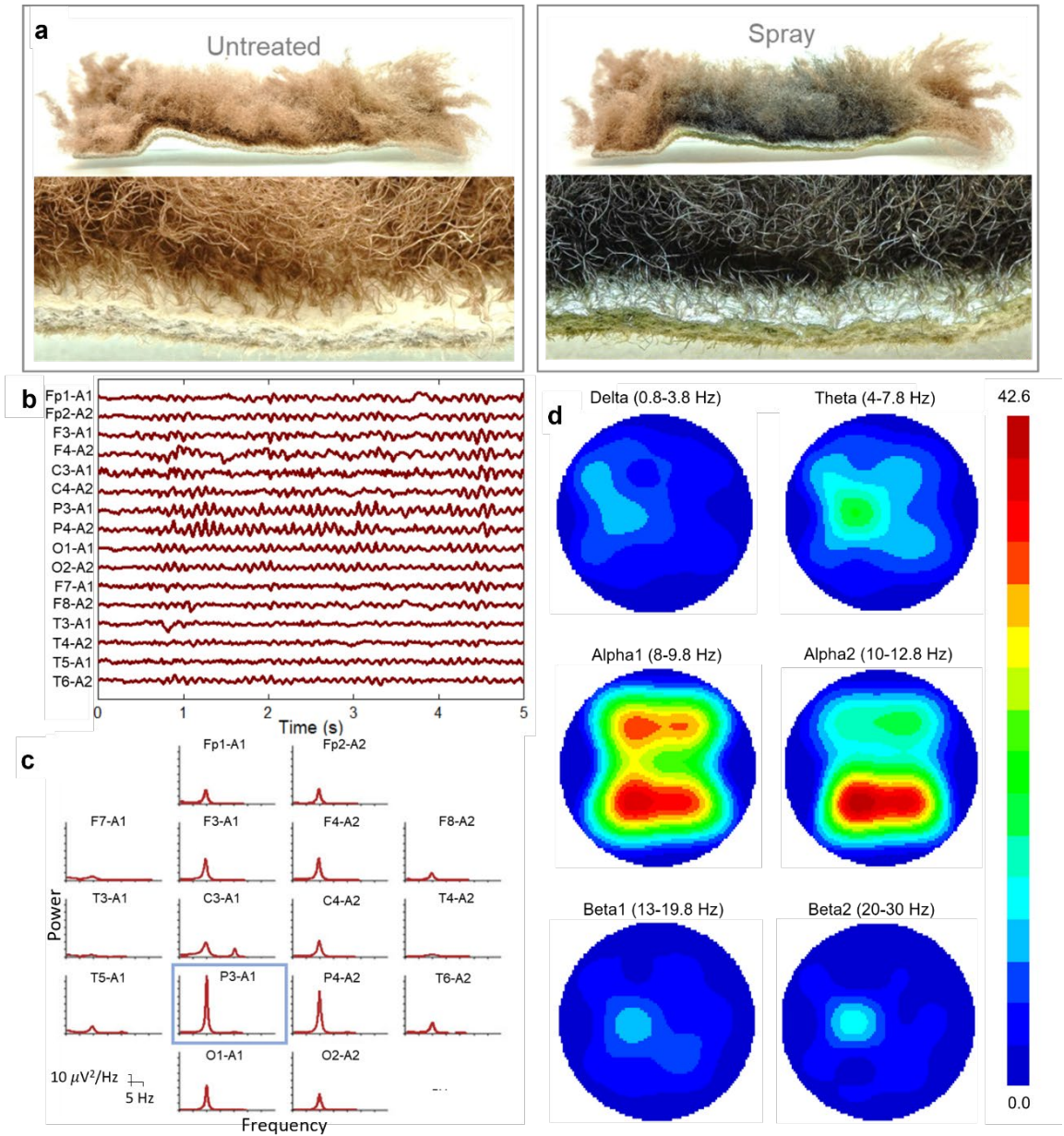

**Fig. 4| Probing electrical potential from functionalized unshaved scalp. a**, Photo of cowhide before and after the VDWTF-EFBS implementation, visualizing the surface morphology of the functionalized hairy body surfaces. After the spray, the fully conformal 2D coating turns the hair dark gray. The skin turns silver due to enhanced reflectance. **b**, Raw EEG waveforms extracted with VDWTF-EFBS when the eyes are closed. Clear alpha waves are observable from the plot. **c**, The Fourier-transformed spectra at different positions. **d**, Spatial distribution of different EEG bands. Nose facing up.

**Bioimpedance changes deep in unshaved scalps**

We further used VDWTF-EFBS to measure the scalp AC impedance. Scalp impedance offers a convenient non-invasive measure of deep-scalp activities[26,44] with better mobility and portability than other methods, such as functional magnetic resonance imaging (fMRI) that requires a strong magnetic field. Hence, it can potentially enable wearable strategy for continuous tracking deep-scalp activities over routine nonclinical activities. We chose two far-separated positions, Fp1 and O2 (by the EEG convention in Supplementary Fig. 6c), to inject the 50 kHz AC current following the recommendation by previous EIT works[45], and used two electrodes at T5 and O1 to measure the voltage (Fig. 5a). Our measurement shows only 0.01% fluctuations over the total signal. In comparison, results with commercial electrodes exhibit random drifts with a standard deviation of 0.32% of the total signal (Supplementary Fig. 9).

Multiple conductive pathways contribute to the measured impedance signal (Fig. 5b). Previous studies suggested that blood dominates brain conductance. Moreover, the saliva in the mouth is also a conductive fluid. Muscle and skin movements, though less pronounced on the scalp region, could also contribute. To this end, we measure waveforms during frequent eye blinking, regular breathing, and swallowing saliva to explore the contribution of different conduction paths (Fig. 5c). Eye blinking does not provide observable signals. Breathing causes small impedance fluctuations with the signature periodicity of a few seconds. Saliva swallowing (gray shades in Fig. 5c) provides a larger spike, followed by a gradual relaxation back to the normal value. Interestingly, the spike amplitude is proportional to the interval between two swallows. Since the muscle movements cease fast after swallowing and are not related to the swallowing interval, the impedance spikes are attributed to saliva conduction. The longer the interval between two swallows, the larger the volume of saliva is secreted, thus leading to a larger percentage change of the total impedance by each swallowing.

Next, we investigated the contribution of blood volume change to the scalp impedance. We first modulated the brain blood volume by having the subject repeatedly stand up and lie down, with each gesture held for 2 min (Fig. 5d). The scalp impedance notably drops when the subject lies down, forcing more blood into the scalp. After standing up, the blood quickly redistributes to the body, causing orthostatic hypotension and an instant increase of the scalp impedance. We evaluated the signal difference between the up ($S_{up}$) and down ($S_{down}$) in six continuous exercise sequences over three different days on two test subjects (Supplementary Fig. 8). The signal change percentage $(S_{up} - S_{down})/S_{up}$ (Fig. 5d inset) reveals an initial signal increase between the first two cycles, after which the absolute difference of the impedance gradually decreases, **indicating that the body gradually adapts to the exercise and initiates a cerebral blood flow autoregulation**[46-48] to stabilize the brain blood volume against the drastic movements. We further tracked the arm arterial pulse wave with a commercial cuff device (Fig. 5e), which showed little adaptation, suggesting the results in Fig. 5d as an intrinsic cerebral-region adaptation, instead of a general cardiovascular response (see also Fig. 2 in ref. 46). The test confirms the correlation between brain blood volume and the measured signals, demonstrating a potentially portable technology for continuously monitoring the brain blood flow during daily activities, which is critical for further blood regulation studies.

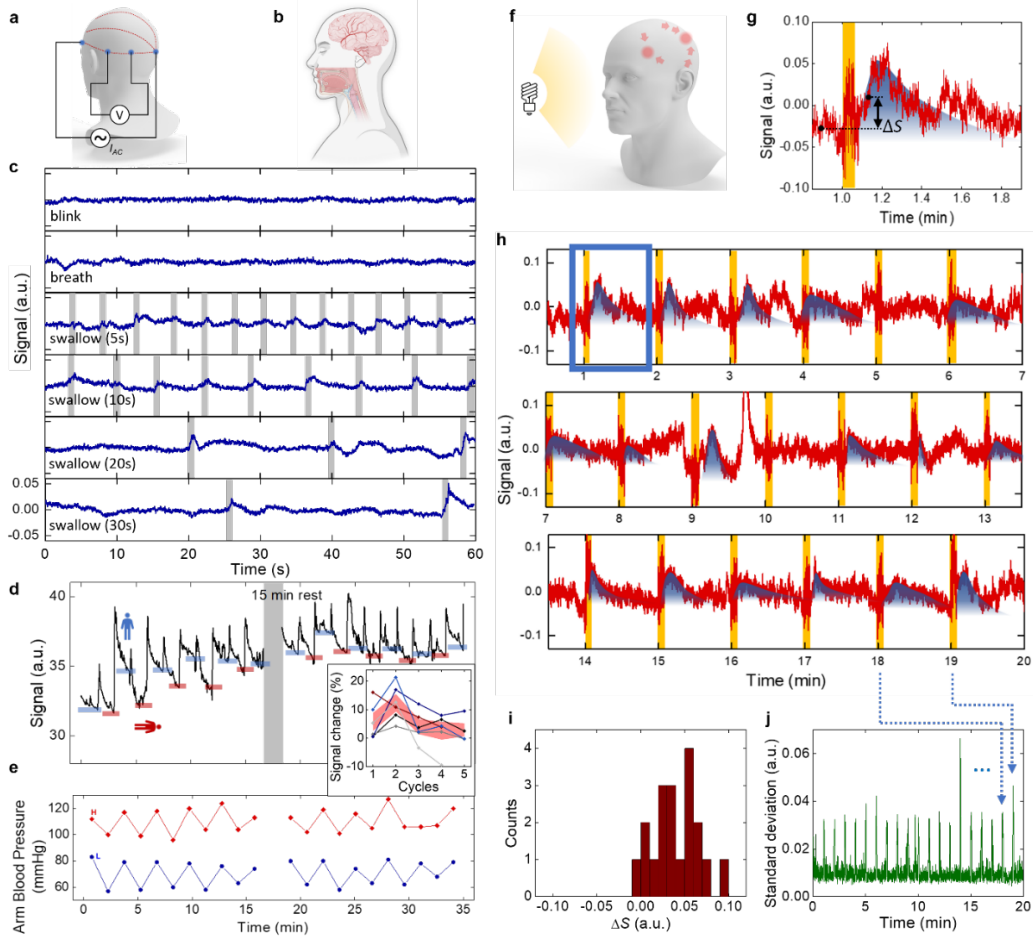

**Fig. 5| Deep-scalp impedance measurements. a**, The measurement profile and the equivalent circuit, with the electric fields indicated by red dashed lines. **b**, Schematic of the major contributors to the measured scalp impedance change, i.e., the neural activities and blood flow in the brain and the saliva in the mouth. **c**, Investigation of activities that contribute to scalp impedance change. The swallowing actions are marked by gray shades. **d**, Scalp impedance measurements during repeated standing up and lying down, with each position holding for 2 min per cycle. After initial motion-related spikes (also observed in previous works[49]), the signal converges to plateaus, which are marked by the colored bars (blue for standing and red for lying). A 15-min rest (shaded gray) was made before the next test to explore the memory effect of cerebral blood flow autoregulation. Inset: statistics of the signal change $(S_{up} - S_{down})/S_{up}$ in six up-and-down tests between each rest cycle. The red shade plots the mean and standard deviation, suggesting an initial increase followed by a subsequent gradual decrease. **e**, The blood pressure measured using a commercial cuffed blood pressure monitor on the left arm during the measurement. The pressure decreases when lying down (with the left arm held on the top side) and increases when standing up. The diastolic pressure (blue) showed slight adaptation behavior, while the systolic pressure (red) showed less changes, indicating adaptation is uniquely related to the cerebral region. **f**, Schematic of visual stimulation using a white LED, triggering neural activity and brain blood redistribution. **g**, Waveforms with eyes illuminated (highlighted in yellow) for 5 s. The observed slower impedance changes are marked by navy shades. **h**, 19 repeated cycles of visual stimulations in 20 min. Each stimulation is followed by 1-min rest to minimize possible memory effects. The blue box marks the waveform in **g**. **i**, The net signal increases from 5 s before illumination to 5 s after illumination shut off ($\Delta S$ in **g**), evaluated from 19 exposures in **h**. The distribution centers above zero, suggesting repeatable signal increase even after the light is shut off, pointing to slower brain blood flow dynamics. **j**, The standard deviation of signals in 1-second time window evaluated from **h**, showing notably larger oscillation synchronized with the illumination. The blue arrows connect the corresponding peaks to the raw signal in **h**.

The brain impedance can also provide valuable information on other brain activities, such as epilepsy[50], stroke[51-53], and visual stimulations[54]. For example, we observed brain impedance response to visual stimulations upon a white LED illuminating the subject's eyes for 5 seconds at the illuminance of 5 klux (Fig. 5f), followed by a 1 min rest in darkness, then exposed to illumination again. The measured impedance signal showed drastic oscillation during the illumination (highlighted in yellow in Fig. 5g, h), together with a relatively slow increase (5-10 s, marked by navy shades). After the LED was turned off, the signal slowly recovered in 20-30 sec. The fast oscillation was observed for all 19 repeated tests, while the more subtle slower component was observed in 17 repetitions despite other fluctuations from breath and swallowing (Fig. 5h). No patterns were measured when illuminating from the back of the subject, ruling out trivial optoelectronic responses from the setup.

The slow and fast responses were highly repeatable, confirmed by further statistics. We sampled the signal 5 s before illumination and the signal 5 s after the same illumination is shut off and computed the difference ($\Delta S$ in Fig. 5g). $\Delta S$ distributes above zero (Fig. 5i) for the 19 illuminations, suggesting a general, repeatable signal increase. We also evaluated the fast oscillation amplitude by measuring the standard deviation of the raw signal with a sliding 1-s time window. The oscillation amplitude clearly synchronizes with the illumination (spikes in Fig. 5j) with only one exception from swallowing.

The slower post-illumination response corresponds to slower deep-tissue dynamics, most likely the brain blood consumption and redistribution during and after the visual stimulation. Previous work reported similar signals on rabbits under visual stimulation measured by brain EIT[55], with impedance changes within ± 5%. Tests on human beings[54] also showed impedance fluctuations with amplitude and time scale similar to the slow component in our results. The fast oscillation cannot originate from eye/muscle movements, as blink showed negligible signal (Fig. 5c). Instead, it is likely associated with the ion channel opening during neuronal depolarization[56]. **The strong screening of the thick human skull leads to only ~0.1-1% change of the impedance, which limited the temporal resolution and hindered resolving the fast component in the previous works[56,57]. A highly robust low-impedance contact interface enabled by our VDWTF-EFBS is essential for clearly and consistently resolving such signals.**

**Conclusion**

By formulating and directly spray-coating 2D $MoS_2$ nanosheets on the natural skin and unshaved scalp under bio-friendly conditions, we created electrically functionalized body surfaces as highly robust electrical-body interfaces. The microscopic conformality and dynamic adaptability of VDWTF-EFBS effectively suppress extrinsic contact-induced artifacts, enabling us to probe the intrinsic deep tissue electrophysiological activities with greatly improved SNR. AC bioimpedance measurements using the VDWTF-EFBS capture radial artery pulses and neck tissue movements during speech. When the EFBS was applied to the unshaved scalp, a higher SNR was observed in EEG measurements. Further scalp bioimpedance measurements unravel deep-tissue activity patterns such as saliva secretion/swallowing, cerebral blood flow change during exercise, cerebral blood flow autoregulation, and visual stimulation response with enhanced SNR and temporal resolution. The results demonstrate a powerful technology for universally functionalizing irregular, dynamically moving body surfaces into low-artifact electrical interfaces for continuously tracking deep-tissue electrophysiological activities.

## References

1. Mahmood, M. et al., Fully portable and wireless universal brain–machine interfaces enabled by flexible scalp electronics and deep learning algorithm. *Nat. Mach. Intell.* **1**, 412-422, (2019).

2. Victorino, J. A. et al., Imbalances in regional lung ventilation: a validation study on electrical impedance tomography. *Am. J. Respir. Crit. Care Med.* **169**, 791-800, (2004).

3. Isaacson, D. et al. Imaging cardiac activity by the D-bar method for electrical impedance tomography. *Physio. Meas.* **27**, S43, (2006).

4. Li, X. et al., Investigation of motion artifacts for biopotential measurement in wearable devices. *IEEE BSN* 218-223, (2016).

5. Webster, J. G. Reducing motion artifacts and interference in biopotential recording. *IEEE TBME* **12**, 823-826, (1984).

6. Yin, J. et al., Motion artefact management for soft bioelectronics. *Nat. Rev. Bioeng.* 1-18, (2024).

7. Lee, S., Kruse. J., Biopotential electrode sensors in ECG/EEG/EMG systems, *Analog Dev.* **200**, 1-2, (2008).

8. Lin, M. et al., Soft wearable devices for deep-tissue sensing. *Nat. Rev. Mater.* **7**, 850-869, (2022).

9. Wu, H. et al., Materials, devices, and systems of on-skin electrodes for electrophysiological monitoring and human–machine interfaces. *Adv. Sci.* **8**, 2001938, (2021).

10. Ness, M. J. et al., Neonatal skin care: a concise review. *Int. J. Dermatol*. **52**, 14-22, (2013).

11. Mietzsch, U. et al., Successful reduction in electrode-related pressure ulcers during EEG monitoring in critically ill neonates. *Adv. Neonatal Care* **19**, 262-274, (2019).

12. Yu, Y. et al., Flexible electrochemical bioelectronics: the rise of in situ bioanalysis. *Adv. Mater.* **32**, 1902083, (2020).

13. Cao, J. et al., Anti-friction gold-based stretchable electronics enabled by interfacial diffusion-induced cohesion. *Nat. Commun*. **15**, 1116, (2024).

14. Zhang, Z. et al., A 10-micrometer-thick nanomesh-reinforced gas-permeable hydrogel skin sensor for long-term electrophysiological monitoring. *Sci. Adv*. **10**, eadj5389, (2024).

15. Li, N. et al., Bioadhesive polymer semiconductors and transistors for intimate biointerfaces. *Science* **381**, 686-693, (2023).

16. Luo, Y. et al., Technology roadmap for flexible sensors. *ACS Nano* **17**, 5211-5295, (2023).

17. Jiang, Y. et al., A universal interface for plug-and-play assembly of stretchable devices. *Nature* **614**, 456-462, (2023).

18. Li, Q. et al., Highly thermal-wet comfortable and conformal silk-based electrodes for on-skin sensors with sweat tolerance. *ACS Nano* **15**, 9955-9966, (2021).

19. Wang, Y. et al., A durable nanomesh on-skin strain gauge for natural skin motion monitoring with minimum mechanical constraints. *Sci. Adv*. **6**, eabb7043, (2020).

20. Fang, Y. et al., Solution-processed submicron free-standing, conformal, transparent, breathable epidermal electrodes. *ACS Appl. Mater. Interfaces* **12**, 23689-23696, (2020).

21. Kabiri, A. S. et al., Graphene electronic tattoo sensors. *ACS nano* **11**, 7634-7641, (2017).

22. Kim, Y. et al., Chip-less wireless electronic skins by remote epitaxial freestanding compound semiconductors. *Science* **377**, 859-864, (2022).

23. Kim, K. K. et al., A substrate-less nanomesh receptor with meta-learning for rapid hand task recognition. *Nat. Electron* **6**, 64-75, (2023).

24. Li, J. et al., A tissue-like neurotransmitter sensor for the brain and gut. *Nature* **606**, 94-101, (2022).

25. Yan, Z. et al., Highly stretchable van der Waals thin films for adaptable and breathable electronic membranes. *Science* **375**, 852-859, (2022).

26. Miyamoto, A. et al., Inflammation-free, gas-permeable, lightweight, stretchable on-skin electronics with nanomeshes. *Nat. Nanotechnol.* **12**, 907-913, (2017).

27. Lin, Z. et al., Solution-processable 2D semiconductors for high-performance large-area electronics. *Nature* **562**, 254-258, (2018).

28. Zhang, D. et al., Broadband nonlinear modulation of incoherent light using a transparent optoelectronic neuron array. *Nat. Commun*. **15**, 2433, (2024).

29. Chen, X. et al., CVD-grown monolayer MoS2 in bioabsorbable electronics and biosensors. *Nat. Commun.* **9**, 1690, (2018).

30. Xu, Z. et al., A critical review on the applications and potential risks of emerging MoS2 nanomaterials. *J. Hazard. Mater*. **399**, 123057, (2020).

31. Notman, R., Anwar, J., Breaching the skin barrier—Insights from molecular simulation of model membranes. *Adv. Drug Deliv. Rev*. **65**, 237-250, (2013).

32. Zhou, X. et al., Remote induction of cell autophagy by 2D MoS2 nanosheets via perturbing cell surface receptors and mTOR pathway from outside of cells. *ACS Appl. Mater. Interfaces* **11**, 6829-6839, (2019).

33. Pan, L. et al., A compliant ionic adhesive electrode with ultralow bioelectronic impedance. *Adv. Mater.* **32**, 2003723, (2020).

34. Pan, L. et al., Enhancing Prosthetic Control through High-Fidelity Myoelectric Mapping with Molecular Anchoring Technology. *Adv. Mater.* **35**, 2301290, (2023).

35. Yang, L. et al., Insight into the contact impedance between the electrode and the skin surface for electrophysical recordings. *ACS Omega*, **7**, 13906-13912, (2022).

36. Ershad, F. et al., Ultra-conformal drawn-on-skin electronics for multifunctional motion artifact-free sensing and point-of-care treatment. *Nat. Commun*. **11**, 3823, (2020).

37. Huang, Y. C. et al., Sensitive pressure sensors based on conductive microstructured air-gap gates and two-dimensional semiconductor transistors, *Nat. Electron*. **3**, 59-69, (2020).

38. Zhao, C. et al., Ultrathin Mo2S3 nanowire network for high-sensitivity breathable piezoresistive electronic skins. *ACS Nano* **17**, 4862-4870, (2023).

39. Foster, K. R., Lukaski, H. C., Whole-body impedance--what does it measure? *Am. J. Clin. Nutrition* **64**, 388S-396S, (1996).

40. Chang, C. C. et al., Electrical impedance tomography for non-invasive identification of fatty liver infiltrate in overweight individuals. *Sci. Rep*. **11**, 19859, (2021).

41. Pesti, K. et al., Electrode placement strategies for the measurement of radial artery bioimpedance: Simulations and experiments. *IEEE Trans. Instrum. Meas.* **70**, 1-10, (2020).

42. Kireev, D. et al., Continuous cuffless monitoring of arterial blood pressure via graphene bioimpedance tattoos. *Nat. Nanotechnol.* **17**, 864-870, (2022).

43. Li, J. et al., Thin, soft, wearable system for continuous wireless monitoring of artery blood pressure. *Nat. Commun*. **14**, 5009, (2023).

44. Holder, D. S., Detection of cerebral ischaemia in the anaesthetised rat by impedance measurement with scalp electrodes: implications for non-invasive imaging of stroke by electrical impedance tomography. *Clin. Phys. and Physiol. Meas*. **13**, 63, (1992).

45. Electrical impedance tomography: methods, history and applications. *CRC Press*, (2004).

46. Claassen, J. A. H. R. et al., Regulation of cerebral blood flow in humans: physiology and clinical implications of autoregulation. *Physiol. Rev*. **101**, 1487-1559, (2021).

47. Koep, J. L. et al., Autonomic control of cerebral blood flow: fundamental comparisons between peripheral and cerebrovascular circulations in humans. *J. Physiol*. **600**, 15-39, (2022).

48. Querido, J. S., Sheel, A. W., Regulation of cerebral blood flow during exercise. *Sports Med*. **37**, 765-782, (2007).

49. Ashley, J. et al., Cerebral blood flow dynamics: Is there more to the story at exercise onset? *Physiol. Rep*. **11**, e15735, (2023).

50. Witkowska-Wrobel, A. et al., Feasibility of imaging epileptic seizure onset with EIT and depth electrodes. *NeuroImage* **173**, 311-321, (2018).

51. Romsauerova, A. et al., Multi-frequency electrical impedance tomography (EIT) of the adult human head: initial findings in brain tumours, arteriovenous malformations and chronic stroke, development of an analysis method and calibration. *Physiol. Meas.* **27**, S147, (2006).

52. McDermott, B. et al., Bi-frequency symmetry difference EIT—Feasibility and limitations of application to stroke diagnosis. *IEEE J. Biomed. Heal. Info*. **24**, 2407-2419, (2019).

53. Yang, L. et al., A novel multi-frequency electrical impedance tomography spectral imaging algorithm for early stroke detection. *Physiol. Meas*. **37**, 2317, (2016).

54. Tidswell, T. et al., Three-dimensional electrical impedance tomography of human brain activity. *NeuroImage* **13**, 283-294, (2001).

55. Holder, D. S., Rao, A., Hanquan, Y., Imaging of physiologically evoked responses by electrical impedance tomography with cortical electrodes in the anaesthetized rabbit. *Physiol. Meas*. **17**, A179, (1996).

56. Gilad, O., Holder, D.S., Impedance changes recorded with scalp electrodes during visual evoked responses: implications for electrical impedance tomography of fast neural activity. *NeuroImage* **47**, 514-522, (2009).

57. McCann, H. et al., Sub-second functional imaging by electrical impedance tomography. *IEEE EMBC* 4269-4272, (2006).

## Methods

**VDWTF-EFBS implementation.** The MoS2 ink was formulated in isopropanol (IPA) following the procedure reported in the previous work[28]. The ink was put into a spray gun and blown to the targeted area with nitrogen. Once sprayed, the IPA quickly evaporates, leaving the 2D conformal nano-coatings on the body surface. We used tapes and patterned PDMS films as the shadow mask to control the EFBS pattern. It should be noted that the spray step was performed in a hooded area with personal protection equipment (facial masks and goggles for neck and head region) to avoid possible exposure to nanoparticles suspended in the air. We kept spraying until the color of the surface turned from green to silver gray, indicating that the thin film was uniform and above the optical absorption depth (tens of nanometers) of the $MoS_2$ thin film. We used a multimeter to measure the lateral resistance of the VDWTF at the two ends of the patterned stripe, and stopped spraying when the readout is below 100 k$\Omega$, so as to control the repeatability of the on-skin thin film property. Further control is possible with an automated spray machine if implemented as a product in the future. We also note that the skin surface of different individuals still introduces variations even with state-of-the-art implementation control, which is the case for all wearable electronics. The issue can be further alleviated with calibrations after implementation. The graphite syrup was made by dispersing graphite powers to commercially available Coke (Coca-Cola, original flavor) drink at 0.1 g/mL. The solution was then heated up to 90 ℃ and concentrated to increase the viscosity and conductivity until a water concentration of 56% calibrated by weight. The cooled syrup was dropped on the contact area. For surface impedance measurements, unconcentrated Coke syrup was mixed with graphite powder and NaCl at a mass ratio of 1:1:1 to further reduce the series resistance extrinsic to the $MoS_2$-skin interface below 0.5 k$\Omega \cdot cm^2$ measured at a thickness of 0.3 mm. The gold-on-SEBS film bridges the VDWTF-EFBS with the aluminum tape electrode for on-arm and on-neck measurements. We dissolved 1 g SEBS in 10 mL toluene and drop-casted the mixture on glass slides. The solution flows and forms a thin film after the toluene evaporates. For safety considerations, we left the thin film in a fume hood for an hour to completely remove the toluene. 40 nm of gold was evaporated on the SEBS layer at a deposition rate of 1 angstrom/s using an electron beam evaporator. The film can be peeled off the substrate in water and dried for subsequent transfer. A layer of scotch tape was stuck between the skin and the aluminum tape for electrical insulation. For EEG measurements on the unshaved scalp, a commercial headset (KT88, Contec Medical Systems Co., Ltd) was used.

**Contact conformality analysis.** Here, we use mechanical modeling to quantify the microscale/nanoscale conformality of the VDWTF EFBS and its roles in reducing contact impedance and motion artifacts. Human skin takes a highly random surface morphology, with surface curvature distribution from micrometers near the trenches to tens of micrometers at the upper surface of stratum corneum cells. We first simplify the skin surface as a sinusoidal function in 1D, with the winkle periodicity of $\lambda$. As shown in Supplementary Fig. 1, a free-standing elastic membrane can either partially or fully conform with the surface due to the competition of the different energy components, including the shear bending energy of the membrane ($U_{bending}$), the energy when the membrane is strained to conform to the skin ($U_{membrane}$), the surface bonding (adhesion) energy ($U_{adhesion}$) and the substrate strain energy ($U_{substrate}$) when it experiences additional strain from the membrane.

$$U_{total} = U_{bending} + U_{membrane} + U_{adhesion} + U_{substrate} \quad (1)$$

The equilibrium state morphology should minimize $U_{total}$.

Prior work[58] reveals that $U_{total}$ can be normalized and reduced as a function of four separate parameters, $\alpha = E_m/E_s$, $\beta = 2\pi h_0/\lambda$, $\eta = t/\lambda$, and $\mu = \gamma/(E_m \lambda)$, with the related parameters illustrated in Supplementary Fig. 1.

$$\widehat{U}_{total} = \frac{1}{12}\alpha\xi^2\eta^3 D(\hat{x}_c) + \alpha\eta\beta^2\xi^4 K(\hat{x}_c) - \frac{\mu}{\beta^2}E(\hat{x}_c) + \frac{(1-\xi)^2}{4\pi}[F_1(\hat{x}_c) - F_2(\hat{x}_c)] \quad (2)$$

Where $\hat{x}_c = 2x_c/\lambda$ is the normalized degree of conformability, and $\xi = h_1/h_0$, $h_1$ is the new equilibrium winkle amplitude due to additional membrane strain effects. The remaining *D, K, E,* and *F* functions are defined by:

$$D(\hat{x}_c) = \frac{2}{1-\hat{x}_c}\sin^2(\pi\hat{x}_c) + \pi^2\hat{x}_c + \frac{\pi}{2}\sin(2\pi\hat{x}_c) \quad (3)$$

$$K(\hat{x}_c) = \frac{\beta^2}{107520\pi}[96\pi(\hat{x}_c - 1)(5\beta^2\cos(2\pi\hat{x}_c) - 5\beta^2 - 28)\sin^4(\pi\hat{x}_c) + 35(144\pi\hat{x}_c + 60\beta^2\pi\hat{x}_c - (96 + 45\beta^2)\sin(2\pi\hat{x}_c) + (12 + 9\beta^2)\sin(4\pi\hat{x}_c) - \beta^2\sin(6\pi\hat{x}_c))] \quad (4)$$

$$E(\hat{x}_c) = \hat{x}_c\left(1 + \frac{\beta^2\xi^2}{4\pi}\right) - \frac{\beta^2\xi^2}{8\pi}\sin(2\pi\hat{x}_c) \quad (5)$$

$$F_1(\hat{x}_c) = \frac{1}{\lambda}\int_0^{x_c} 2\left(1 + \cos\frac{2\pi x}{\lambda}\right)\cos\left(\frac{\pi x}{\lambda}\right)\sqrt{\sin^2\frac{\pi x_c}{\lambda} - \sin^2\frac{\pi x}{\lambda}}\, dx \quad (6)$$

$$F_2(\hat{x}_c) = \frac{1}{\lambda}\int_0^{x_c}\left(1 + \cos\frac{2\pi x}{\lambda}\right)\sin^2\left(\frac{\pi x}{\lambda}\right)\cos\left(\frac{\pi x}{\lambda}\right)/\sqrt{\sin^2\frac{\pi x_c}{\lambda} - \sin^2\frac{\pi x}{\lambda}}\, dx \quad (7)$$

We first use the typical values of PDMS-skin contacts as an example. The skin Young's modulus is $E_s = 130\ kPa$. The Young's modulus of PDMS is $E_m = 4\ MPa$. The adhesion coefficient is estimated[59] to be 0.05 J/m². The skin roughness is set as $h_0 \sim 5\ \mu m$, typical for reported skin roughness. The winkle periodicity is set to be $\lambda = 50\ \mu m$. The PDMS thickness is $t = 1, 2, 5, and\ 10\ \mu$m. We can calculate the energy landscapes as shown in Supplementary Fig. 2. The minimum energy gives the actual conformality ($\hat{x}_c \times 100\%$) and the modified wrinkle roughness ($h_1$). Consistent with previous experimental and theoretical studies, a thicker membrane carries more elastic energies, including the bending and strain energies, and leads to poorer conformality.

We further simulate different membranes, including VDWTF, 1 mm ion gel for commercial Ag/AgCl gel electrodes, 1 $\mu m$ PDMS film, and 100 nm free-standing gold film. For VDWTF, the previous works suggested deep anchoring into the lipid bilayers, enabling ~1000 kJ/mol bonding energy[31,32] compared with ~4 kJ/mol. Taking a conservative estimation of 100-time adhesion enhancement, $\gamma = 5$ J/m for VDWTF. The Young's modulus of VDWTF was reported in our previous work to be 47.3 MPa. We previously achieved continuous thin films as thin as 10 nm at the sheet conductance below 100 kΩ. Hence, a 30 nm thick film can easily be achieved for VDWTF EFBS with decent conductivity. The results (Supplementary Fig. 3a) show conformality down to

3 μm, with a lower radius of curvature less relevant to skin morphology. The commercial gel electrodes start to fail at hundreds of micrometers. Considering the stratum corneum cells with diameters around 30 μm, which randomly pack to form the outer skin surface, there are lots of skin features smaller than the scale that gel electrodes fail to conform. Other work[33] also observed air bubbles at the interface surrounded by viscous gel at the skin interface, which further reduced the contact area and led to orders of magnitude larger interface impedance. Thin PDMS performs better than gel electrodes, which has been demonstrated by many previous works. The gold film performs worse than PDMS even with a 10-time thinner membrane because of the large Young's modulus.

Next, we expand the discussion to the impact of conformality on motion artifacts. Assuming the motion generates a strain $d\varepsilon$, then $\lambda$ expands to $(1+d\varepsilon)\lambda$, and lead to a conformal area percentage change $\frac{da}{a} = \frac{dA}{A}$, $A$ is the total conformal surface area. Hence, the interface resistance and capacitance also change by the same amount: $-\frac{dR_I}{R_I} = \frac{dC_I}{C_I} = \frac{da}{a}$. The motion artifact, defined as the contact impedance fluctuation due to the strain, is $\frac{1}{Z}\frac{dZ}{d\varepsilon} = -\frac{1}{a}\frac{da}{d\varepsilon} = -\frac{\lambda}{a}\frac{da}{d\lambda}$, which can be numerically derived from Supplementary Fig. 3a. The result is plotted in Supplementary Fig. 3b. The partial conformity, even at high conformity percentages, can introduce severe motion artifacts because of the large slope $\frac{da}{d\lambda}$ at partial conformity, even if it is near 100%.

Consequently, the theoretical modeling and the numerical simulation suggests that the VDWTF EFBS considerably reduce the contact impedance, and more importantly, reduces the motion artifact induced by contact area change during motions, which is the underlying physics that enables the experimental work.

**Four-probe AC impedance measurement.** The AC current injection was enabled with the internal trigger of the SR850 lock-in amplifier. To ensure contact-robust four-probe measurement, we convert it to a good constant voltage source by attaching a large resistor at the signal output. The impedance of the resistor at the operation frequency was calibrated to be 0.42 MΩ. We applied a voltage of 5 Vpp, injecting a current of 12 μApp, two orders of magnitude lower than the typical safe level. The AC frequency was chosen to be 90 kHz, close to the upper frequency limit of the lock-in amplifier, to achieve lower skin impedance with respect to the artery impedance, which improves the SNR.

**SNR evaluations.** Multiple bioactivity characterizations were involved in the experiments (see Supplementary Table 1) with different SNR evaluation protocols. For DC resistance tests in Fig. 2, the motor has a high rotation speed of around 1000 rpm, hence generating an oscillation above 10 Hz. The natural fluctuation of body resistance, which is used as our signal, is much slower and peaks at 1-2 Hz (Supplementary Fig. 4). Consequently, we choose 3 Hz as the frequency threshold to separate the signal and noise of the body resistance. The SNR is evaluated by the spectral power ratio between the low-frequency band and the high-frequency band, following the equation below:

$$SNR = 10 \cdot log_{10} \frac{\sum_{i=1}^{N} x_i^2}{\sum_{i=1}^{N}(s_i - x_i)^2}$$

$x_i$ is the signal determined by the frequency. $s_i$ is the total raw signal. The summations are run on all sampling points.

The same spectral power-determined SNR was also evaluated for EEG measurements on unshaved scalp, however with different frequencies considered (for example, evaluation of alpha waves used the 8-12 Hz band as the signal, and the other bands as the noise) following a previously reported protocol that was used to evaluate commercial EEG headsets[60].

For ECG, because there are time intervals that minimal signal power presents, we segmented the signal in the time domain, then evaluated the peak signal power (fist making and ECG spikes) as the signal. The noise power at the flat sections is calculated to provide the SNR.

For AC body impedance measurements on the wrist, neck, and unshaved scalp, the signal peaks (base frequency, second harmonic, and third harmonic) are below 5 Hz (Supplementary Fig. 5c). Hence, we used the data below 5 Hz as the signal and oscillations above 5 Hz as the noise. Data with a 10-s time duration was used to calculate the spectra. Note that an additional 0.5 Hz high pass filter was applied for blood pressure SNR evaluations, since the waveform, instead of the background, is important for the particular evaluation.

**Blood pulse waveform analysis.** The P-peak pulse width was evaluated as the full-width-half-magnitude (FWHM) of the peak counted from $U = 0$, see Supplementary Fig. 5a. The error was defined by the standard deviation of 5 continuous waveforms for both the VDWTF-EFBS and the commercial pads.

**Alphabet retrieval from AC impedance waveforms.** We collected 25 waveforms for each pronounced alphabet, with sufficient time intervals to separate between two phonations. The waveforms were segmented and aligned for each alphabet, forming waveform vectors $v_j$. The footnote $j$ denotes the particular waveform data. Next, we randomly shuffled all the waveforms and their corresponding labels, and segmented the shuffled data into five groups. Four groups were used as the training dataset, and the other group as the test dataset. For each alphabet, we compute the mean vector of the training waveforms for the letter $\alpha$:

$$\overline{v_\alpha} = \frac{1}{N_\alpha} \sum_{\in \alpha} v_j$$

$N_\alpha$ is the number of $v_j$ in the training waveforms that correspond to letter $\alpha$. For a test waveform vector $v_k$, we define the similarity metrics $S_{\alpha k}$ between it and $\overline{v_\alpha}$ as their normalized dot product (i.e., the nearest centroid classification method widely adopted in machine learning):

$$S_{\alpha k} = \frac{\overline{v_\alpha} \cdot v_k}{|\overline{v_\alpha}||v_k|}$$

The predicted letter for $v_k$ is $\beta_k$ with maximized $S_{\alpha k}$ over all letter $\alpha$'s.

$$\beta_k = \max_\alpha S_{\alpha k}$$

The prediction is run over all test data for one round of classification test. To minimize the randomness of accuracy due to the shuffle process, we further used the standard cross-validation procedure, i.e., permute over all the 5 shuffled subgroups as the test data and the remaining groups as the training data. Accuracy is evaluated by averaging all five rounds of classification tests. More advanced algorithms that involve additional nonlinearities could possibly provide even higher classification accuracies.

**Data availability**

All data are available in the manuscript or the supplementary materials.

**Code availability**

A custom nearest-centroid classifier is used for retrieval of vocal alphabets based on the neck impedance signals. The code is deposited in https://github.com/DrRitardo/Skintronics2024

58. Wang, L., Lu, N., Conformability of a thin elastic membrane laminated on a soft substrate with slightly wavy surface. *J. Appl. Mech.* **83**, 041007, (2016).

59. Yu, Y. L. et al., Work of adhesion/separation between soft elastomers of different mixing ratios. *J. Mater. Res*. **30**, 2702–2712, (2015).

60. Radüntz, T., Signal quality evaluation of emerging EEG devices. *Front. Physiol.* **9**, 98, (2018).

**Acknowledgement**

We acknowledge the California NanoSystems Institute (CNSI) at UCLA for the device fabrication and technical support. **Funding:** The project is supported by the UCLA CNSI Noble Family Innovation Fund. X.D. acknowledges partial support from the Office of Naval Research through grant no. N00014-22-1-2631 for 2D device fabrication and optimizations.

**Author contributions**

X.D. conceived of the research. X.D., D.Z., and D.X. designed all experiments. D.Z. and D.X. developed the VDWTF-EFBS implementation process. D.Z., D.X., Y.Z., Y. Ling, and Y. Liu prepared the nanomaterials for the test. S.W., K.W., Q.C., J. Y., E.Z., X.Z., and J.C. contributed to the biomedical characterization designs and assisted with the on-body measurements. D.Z., D.X., Y.Z., and B.Z. performed the electronic measurements. D.Z. analyzed the on-body deep-tissue measurement results. D.Z., Y.Z., D.X., B.Z., and X.D. conducted and participated in the on-body measurements. C.W. took optical images for characterizing the EFBS morphology on different body surfaces. X.D. and D.Z. co-wrote the paper. X.D., Y.H., and T.K.H. supervised the research. All authors discussed the results and commented on the manuscript.

**Competing interests**

A provisional patent application is being filed on the directly-coated stretchable van der Waals thin films. The authors declare no other competing interests.

**Additional Information**

Supplemental information is available for this paper.

**Correspondence and requests for materials** should be addressed to Y.H. and X.D.

**Reprints and permissions information** is available at [*link*]